\documentclass[a4paper,reqno,12pt]{article}
\usepackage[right=1.4in, left=1in]{geometry} 
\usepackage{verbatim}
\usepackage{amsfonts}
\usepackage{amssymb}
\usepackage{amsmath}
\usepackage{amsthm}
\usepackage{color}
\usepackage[utf8]{inputenc}
\usepackage{tikz}
\usetikzlibrary{decorations.pathreplacing,angles,quotes}
\usepackage[T1]{fontenc}
\usepackage{graphicx}
\usepackage{caption}
\usepackage{subcaption}
\usepackage{url}
\usepackage[colorlinks,allcolors=blue]{hyperref}
\usepackage{multirow}
\usepackage{rotating}
\usepackage{tabularx}

\begin{document}
\title{
Clustering of countries based on the associated social contact patterns in epidemiological modelling}

\author{Evans Kiptoo Korir and Zsolt Vizi}

% \address{University of Szeged, Bolyai Institute, \\
% Aradi vértanúk tere 1, \\ 
% 6720, Szeged, Hungary\\ 
% E-mail: zsvizi@math.u-szeged.hu}

\maketitle

\abstract{
Mathematical models have been used to understand the spread patterns of infectious diseases such as Coronavirus Disease 2019 (COVID-19). The transmission component of the models can be modelled in an age-dependent manner via introducing contact matrix for the population, which describes the contact rates between the age groups. Since social contact patterns vary from country to country, we can compare and group the countries using the corresponding contact matrices. 
In this paper, we present a framework for clustering countries based on their contact matrices with respect to an underlying epidemic model. Since the pipeline is generic and modular, we demonstrate its application in a COVID-19 model from \cite{RB} which gives a hint about which countries can be compared in a pandemic situation, when only non-pharmaceutical interventions are available.
}


\section{Introduction}

The spread of respiratory infections such as measles, influenza, and the recent COVID-19 occur predominantly through social contacts between susceptible and infected persons. Quantification of these contacts relevant is significant to predict disease dynamics and effect of introduced strategies that aim to control and curb further spread \cite{LD}. Epidemic models (especially deterministic ones) are widely used to understand and predict the disease dynamics and the impact of new strategies and interventions such as lockdown, school closure or vaccination to control infectious disease transmission. Using these models, individuals are assumed to follow certain age specific contact patterns \cite{KR, MH, PK, RB, WH}. These patterns are often inferred from diary based surveys of self-recorded contacts which captures demographic information (age, household size, gender and occupation) \cite{KK}. Such surveys are rare and limited in nature due to economic cost involved.

The rate of contacts between people is not same as it depends on factors such as individual behaviors, age groups, gender, and location of the contact (home, school, workplace, community or other setting)  \cite{KK, KR, MX, MH, PK, RB}. Age group is a critical determinant of disease transmission, and it is usually represented by contact matrices. The elements of these matrices are the average number of contacts an individual in some age group $i$ makes with other individuals belonging to each age group $j$ within a specified period of time, and they are commonly assortative \cite{KK, KR, MX, MH, PK, RB}. Age-structured models provide an understanding of age groups that are key epidemic drivers and the portion of people who are vulnerable to the disease. For an outbreak of infectious diseases such as COVID-19 and influenza, public health introduces non-pharmaceutical interventions (school closure, contact tracing, social distancing, lockdowns, travel restrictions), or public health interventions such as vaccination and antiviral use to alter the pattern of contacts \cite{MH}.

Multiple studies targeting to estimate social contacts have been published. The most referred study, POLYMOD, estimated contact matrices in 8 European countries based on surveys that involved 7290 participants with reported 97,904 contacts \cite{MH}. Authors in \cite{PC} estimated synthetic contact matrices for 152 countries in 2017 based on \cite{MH}, demographic data, surveys, and other sources. The contact matrices were then updated in 2021 with recent data to include 177 countries with introduction of countries such as Hong Kong and the People’s Republic of China \cite{PK}. Fumanelli et al. \cite{FA} estimated the contact matrices by constructing virtual populations in 26 European countries using detailed census and demographic data, an approach that can be adopted in the absence of specific experimental data. Iozzi et al. in their paper \cite{IT} entitled "little Italy" used agent-based model to compute social contact matrices by simulating individual-based model using demographic and Italian use data. Country specific contact matrices have also been estimated, for instance in Uganda \cite{LD}, Kenya \cite{KT},  Zimbabwe \cite{MD},  India \cite{KG}, China \cite{RJ}, Vietnam \cite{HT}, Peru \cite{GG}, Russia \cite{AE}.

As social contact patterns depends on the region of investigation, there is no evidence on spatial comparison of social contacts relevant for transmission between countries. Countries with unfavorable social structures and large population with risky behaviors are likely to have high transmission rates. Understanding this behavior is critical for timely introducing non-pharmaceutical interventions, monitoring the impacts of such decisions and possible reintroduction of new measures in the event of cases resurgence in the countries \cite{MX}. During the COVID-19 world pandemic, decision makers in countries where an outbreak occurred later in time, monitored the actively intervening countries and analyzed the applied strategies to make efficient decisions with proper timing. In these situations, it is crucial to know, which regions have to be monitored and which practices are recommended to be adapted. For this end, social contact patterns made in different countries are grouped into clusters and then examined to understand which countries have similar mixing patterns. 

Here, we employ hierarchical clustering technique to determine the clusters of countries  based on only the social contact patterns. The approach can be extended to include country characteristics such as gross domestic product (GDP), population density, life expectancy, total fertility rate, mortality rates, and other metrics. Since optimization of learning algorithms are highly affected by dimensionality of the data, we handle this problem with usage of  dimension reduction techniques. Additionally, for ensuring comparibility of the underlying contact matrices, we use epidemic models to normalize these matrices as a preprocessing step before the above mentioned steps.

The structure of the paper is the following: in the second chapter we introduce the underlying data and the different pieces of the proposed framework, namely the concept of an age-structured epidemic model, the dimension reduction to avoid curse of dimensionality and possible approaches for clustering the data points represented by a reduced feature vector. The third section gives a demonstration about how the pipeline works with specific setup of the framework. In the last section we discuss the considerable options for modifying and improving the presented methods.

\section{Methods and Materials}
\subsection{Data}
Prem et al. \cite{PK} updated synthetic contact matrices based on \cite{MH} with the most recent data from sources such as the United Nations Population Division, International Labour Organization,  World Bank databases, demographic data, and surveys. The authors estimated contact patterns in four different locations (home, work, school and other locations) for non-POLYMOD countries and made comparison to prior estimated contact matrices. Contact patterns were then projected for a total of 177 countries. 

We used prior results of this constructed social contact matrices in \cite{PK} but targeting only all the European countries. The social contact matrices are of size $16 \times 16$ representing four settings (Home, School, Work, and Other location) among 16 age groups: 0–4, 5–9, 10–14, $\dots$, 75+. A total of 39 countries in Europe were available for analysis (countries are listed in Appendix).  We considered the complete social contact mixing in the form of a \textit{full social contact matrix}:

\begin{equation*}
     M_{C} = M_{C, H} + M_{C, S} + M_{C, W} + M_{C, O}
\end{equation*}
where $M_{C, X}$ is the social contact matrix with type $X\in\{H, W, S, O\}$ from country $C \in \{1, . . . , 39\}$.

When social contacts are measured empirically, reciprocity is usually reported \cite{KK, KR, MX, MH, PK, RB, WH} since contacts are mutual. We corrected each matrices $M_{C}$ to ensure that the total number of contacts from age group $j$ to $i$ is equal to the total number of contacts from age group $i$ to age group $j$:

\begin{equation}
    M^{(i,j)}_{C} P_{i} = M^{(j, i)}_{C} P_{j} %\Rightarrow \frac{M_{ij}^{(C)}}{M_{ji}^{(C)}} = \frac{P_{j}}{P_{i}}
    \label{eq2}
\end{equation}
where $M^{(i, j)}_{C}$ is the element of $M_C$ in the $i$th row and the $j$th column,
$P_{i}$ and $P_{j}$ represents the number of individuals in age class $i$ and $j$ respectively.

Specifically, we want to work with contact matrices satisfying Eq. (\ref{eq2}), thus we applied symmetrization for the underlying data to ensure the consistency of total number of contacts between two age groups:

\begin{equation*}
     M_{C} = \frac{1}{2 P_{i}}\left(M^{(i, j)}_{C} P_{i} + M^{(j, i)}_{C} P_{j}\right)     
\end{equation*}
where $M_{C}$ is the full contact matrix now being corrected for reciprocity. Its elements are average number of contacts between age group $i$ and $j$ with very strong contacts along the diagonals characterising strong age-assortative mixing (interactions between people of same age group). The off-diagonals captures inter-generational mixing (interactions between people of different age groups, for instance, children and their parents).

\begin{figure}[ht!]
    \centering
    \includegraphics[width=0.19\textwidth]{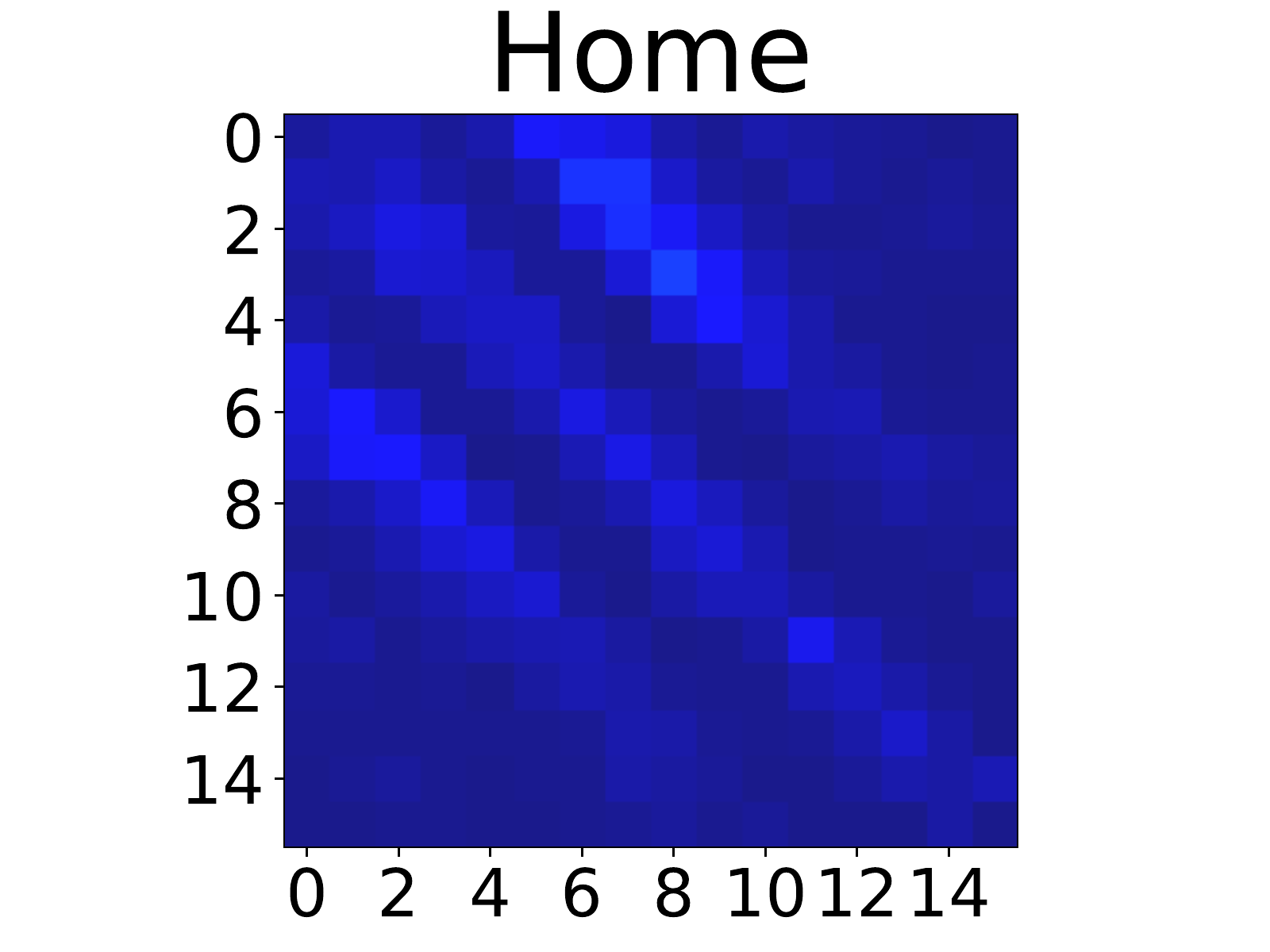}
    \includegraphics[width=0.19\textwidth]{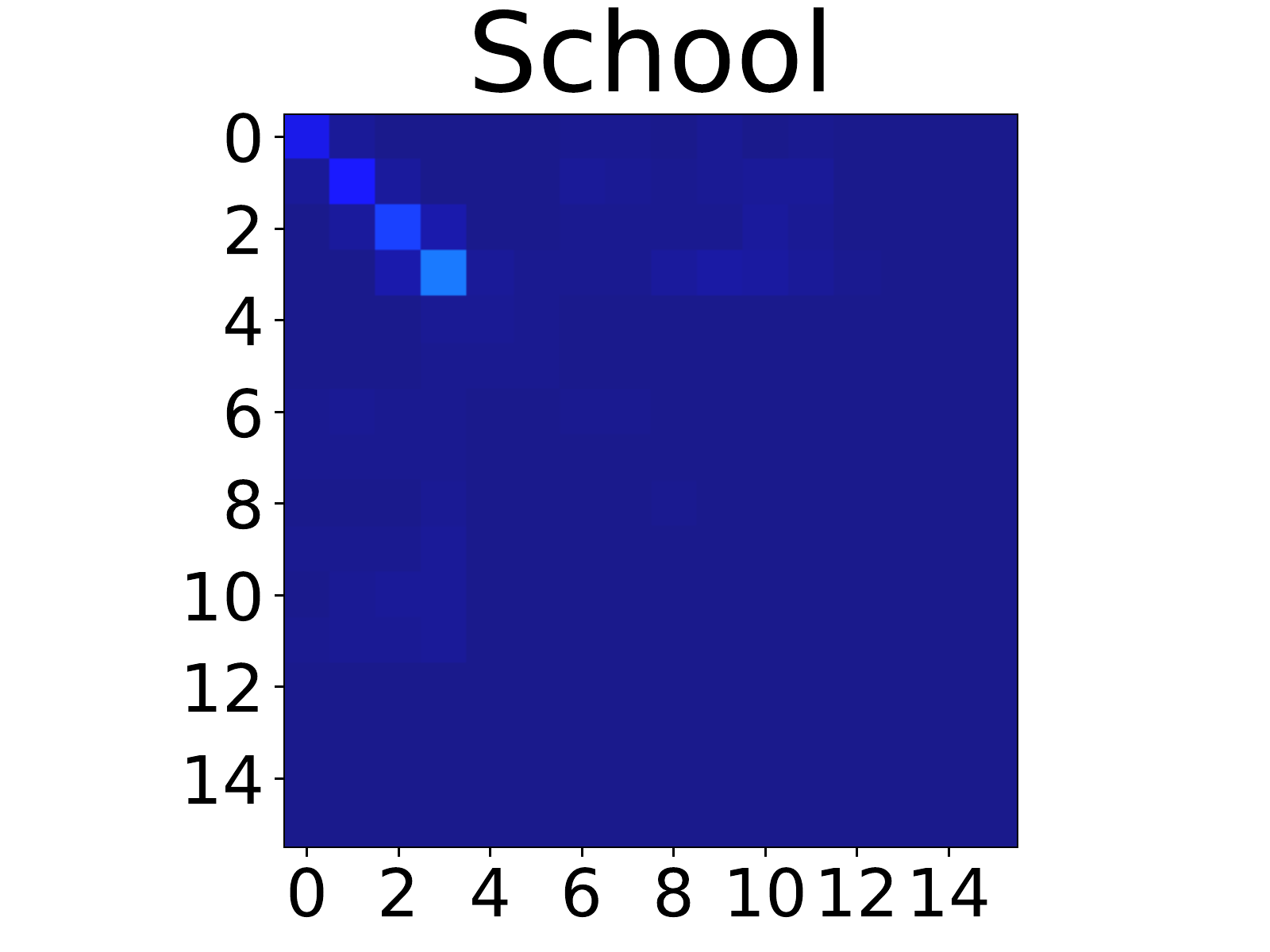}
    \includegraphics[width=0.19\textwidth]{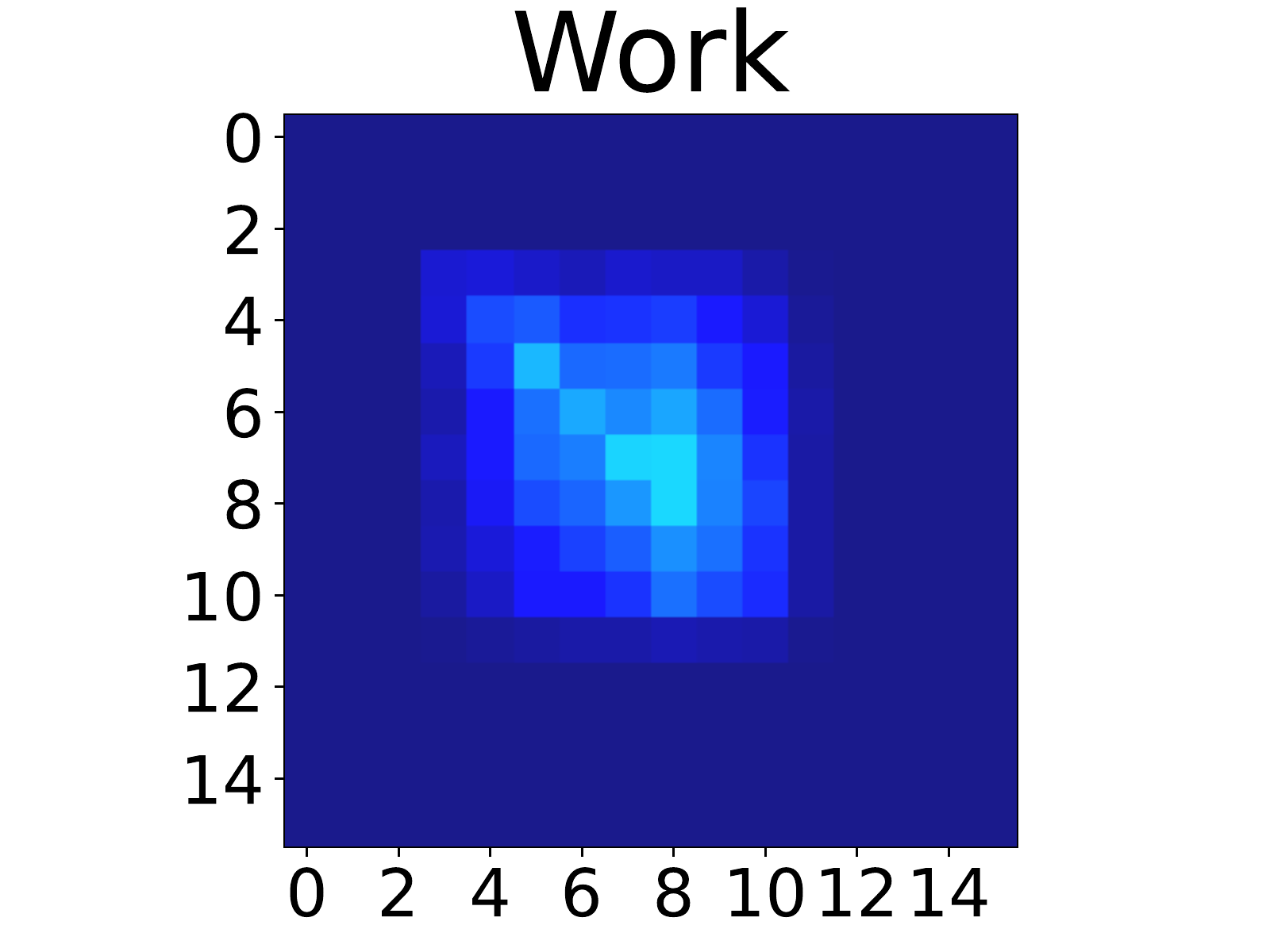}
    \includegraphics[width=0.19\textwidth]{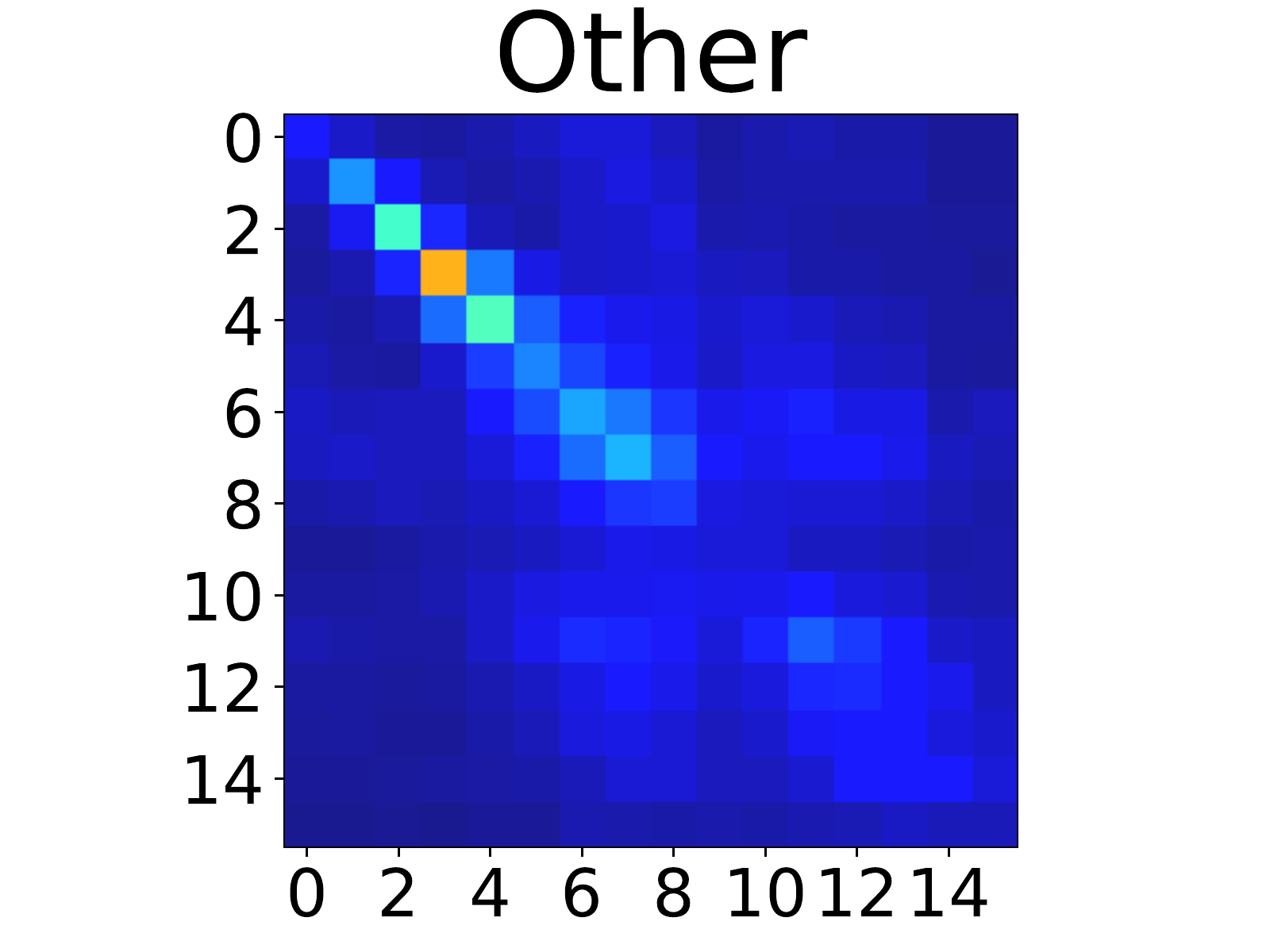}
    \includegraphics[width=0.19\textwidth]{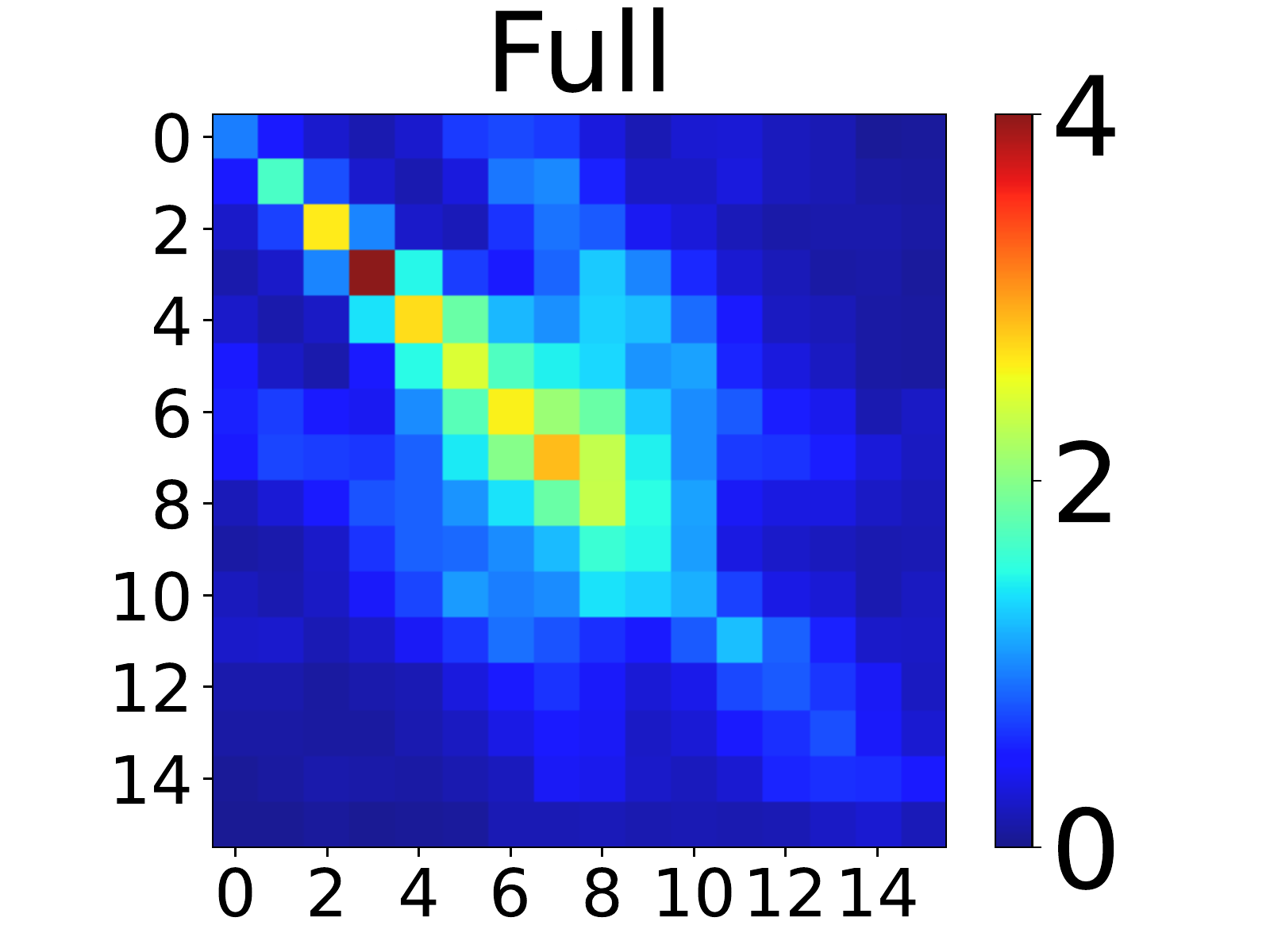}
    \caption{Age specific contact matrices in Hungary. Mixing patterns by age at home, school, work, and other places. The full matrix is obtained as the sum of the matrices represented in the four settings  (home, school, work, and other places). In the horizontal axis, we have the age of the participant that was involved in the survey and vertical axis shows the age of the reported contact. The elements of the matrices are the average number of contacts in the age group of the participant, which increases from dark blue to red on the images}
    \label{fig: hungary}
\end{figure}

Fig. \ref{fig: hungary} shows the contact matrices obtained for Hungary as an example of the social mixing patterns in European countries. The matrices are highly structured and varied considerably between settings. Reported contacts at home in Fig. \ref{fig: hungary} shows a dominant diagonal (assortative pattern) representing contacts with people of the same age group such as young individuals. The off-diagonals reports inter-generational mixing contacts, for instance between children and their parents. Notably, these contacts are least pronounced for people aged over 60. The contacts at school in Fig. \ref{fig: hungary} is evident as young people (people with age 0-19 years) interact mostly with other students of the same age and school. In the workplaces displayed in Fig. \ref{fig: hungary}, we can see that most contacts occur between people in working population age (individuals with age group 20-64 years). These contacts are less assortative than at home. Overall contacts in other places (excludes home, work and school) shows age assortative for younger groups and less assortative for older groups (Fig. \ref{fig: hungary}). The full contacts in Fig. \ref{fig: hungary} which combines all the contacts in the four settings shows strong contacts between people of age 15-59 years in the country.

\subsection{Epidemic Model}

Epidemic models particularly age-structured models provide a better understanding of global spread and enable designing public health interventions. Deterministic compartmental models such as variants of $SIR$, $SEIR$ and $SIS$ are formulated as a system of ordinary differential equations. The proposed framework can consider any age-structured model for analyzing the similarities between countries with different contact matrices.

\begin{figure}[ht]
    \centering
    \includegraphics[width=1\textwidth]{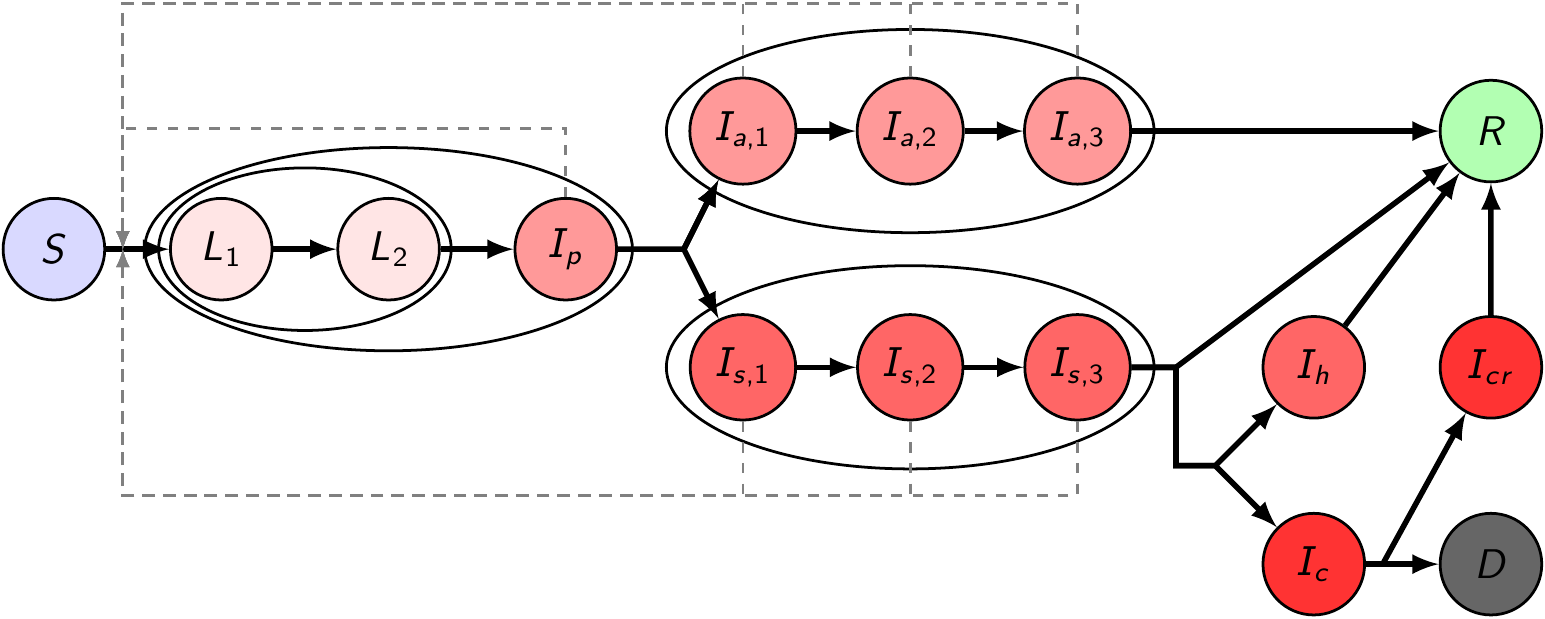}
    \caption{Flowchart of the transmission model from $\cite{RB}$ used for demonstrating the proposed framework.}
    \label{img: trdiagram}
\end{figure}

As a demonstration, we consider the model in Fig.\ref{img: trdiagram} from \cite{RB}. The model has 15 compartments. $S$ denotes the susceptibles, i.e., those who are at the risk of contracting the disease. Latents $(L)$ are infected individuals but do no show symptoms and cannot infect. $I_{p}$ compartment contains the pre-symptomatic individuals who do not show symptoms but are able to infect. Asymptomatic (or mildly symptomatic) and symptomatic infectious individuals, denoted by $I_a$ and $I_s$, respectively, are divided into 3 groups (enabling the modelling of recovery time by an Erlang-distributed variable). Individuals from the the last stage of asymptomatic compartment will all recover and hence proceed to the recovered class $R$ while those in the last stage of symptomatic compartment may recover without the need for hospital treatment (and thus move to $R$) or become hospitalized, either with normal hospital care ($I_h$) or intensive critical care ($I_c$). Those who need normal hospital care will recover and those at the intensive care units may overwhelm the disease proceeding first to a rehabilitation unit ($I_{\mathrm{cr}}$) before complete recovery, or for them fatal outcome might occur ($D$). The example model has an age-structured setup considering heterogeneity of contacts in the population via contact matrices and introduces age-specific parameters for some transitions in the transmission diagram. For the governing equations, see Appendix B. 

In this demonstration, we assume that the model parameters are the same for all countries except the baseline transmission rate, which is strongly related to the corresponding contact matrix, thus we denote by $(\beta_{0, C})$ the baseline transmission rate for country $C$.

An important quantity to calculate in epidemiology is the basic reproduction number $\mathcal{R}_{0}$, which represents the number of secondary infections caused by a single infective introduced into a wholly susceptible population \cite{DH, KK, KR, MX, RB}. For an age-stratified model, the basic reproduction number can be obtained via using the Next Generation Matrix (shortly NGM) method. As a summary (see details in Appendix C), $\mathcal{R}_0$ can be calculated as the dominant eigenvalue of the model-specific NGM and particularly baseline transmission rate can be take out in the computations. Since calculation of the NGM depends on the underlying contact matrix, thus for a fix value of $\mathcal{R}_0$ we can determine the baseline transmission rate $\beta_{0, C}$ for each country $C\in\{1, 2, \dots, 39\}$ (assuming that model parameters and contact matrix are known for the calculations). Using these transmission rate we apply a standardization to all \textit{full} contact matrices enabling comparability of the countries for a fixed value of $\mathcal{R}_0$ and we denote this new matrix by $S_C$:

\begin{equation*}
    S_{C} = \beta_{0, C} \cdot M_{C}.
\end{equation*}

This standardization is used as a normalization for transforming matrices to be distributed on the same interval. Significantly large differences can appear for the matrices depending on how much data was available for the estimation in \cite{PK}. This data preprocessing step is needed for the next, data science-specific parts of the framework.

\subsection{Dimensionality Reduction}
Data with high dimensional input vectors (i.e. order of sample size is at least the order of the dimensionality) pose problems to learning methods, since training performance of these methods is affected by the curse of dimensionality \cite{BB}. This severe problem can be solved by the introduction of dimension reduction technique or feature selection \cite{AK, HF}.

In the classical theory of clustering, the features are arranged into vectors and we often use a projection-based algorithms for reducing the dimensionality of this vector. Principal Component Analysis (shortly PCA) is a popular technique to project the data vectors onto an affine hyperplane. The principal components are the vectors spanning an optimal projection plane and can be computed effectively using covariance matrix or SVD (former solves the optimization problem to maximize explained variance, latter gives the optimal low-rank approximation for the data matrix). We will refer to the classical PCA as 1D PCA.

Matrix-valued data are becoming increasingly common with advancement in the technology, e.g. images in computer vision problems. Driven by 1D PCA, several techniques that aim to preserve the $2D$ structure of the matrix have been developed \cite{WS}. These includes 2D variants of PCA, which consider projection along rows/columns of the matrices and $(\mathrm{2D})^{2}$ PCA, which applies projection along both dimensions. Recent approaches were developed such as the population value decomposition (PVD) that is precisely equivalent to a two-step SVD \cite{ZT}. 

Using $(\mathrm{2D})^{2}$ PCA approach, we concatenate each country-specific full contact matrices to obtain the matrix $A_{\mathrm{col}}=[M_1^\top, M_2^\top, \dots, M_{39}^\top]^\top \in \mathbb{R}^{(39 \cdot 16) \times 16}$. For the demonstration, in the column direction two principal components were retained by projecting $A$ onto the plane spanned by the columns of $Q \in \mathbb{R}^{16 \times 2} $, thus having $$\hat{A}_{\mathrm{col}} = A_{\mathrm{col}} Q \in \mathbb{R}^{(39 \cdot 16) \times 2}.$$
In the next step, along the row direction, we retained two principal components as well and projected the concatenation of the transposed matrices $A_{\mathrm{row}} = [M_{1}, M_{2}, \dots, M_{39}]^\top \in \mathbb{R}^{(39 \cdot 16) \times 16}$ using matrix $P \in \mathbb{R}^{16 \times 2}$ obtaining $$\hat{A}_{\mathrm{row}} = A_{\mathrm{row}} P \in \mathbb{R}^{(39 \cdot 16) \times 2}.$$
Using the two projection matrices $P$ and $Q$, we project all the matrices $M_C$ with size $ 16 \times 16$ applying both $P$ and $Q$ to get the reduced matrix $$\hat{M}_C = Q^\top M_C P \in \mathbb{R}^{2 \times 2},$$ where $C\in\{1, 2, \dots, 39\}$. In the clustering, for country $C$ we use the vector $m_{C}\in\mathbb{R}^4$ calculated by flattening the $\hat{M}_C$ matrix as a feature vector.

Since the contact matrices satisfy Eq. (\ref{eq2}), the lower and upper triangular parts of a contact matrix are not independent from each other. For applying 1D PCA technique, it is beneficial to consider only the independent elements of the contact matrix, thus we will take only the upper triangular part consists of $136$ elements. In summary, the data matrix for 1D PCA will be $A\in\mathbb{R}^{39 \times 136}$, where $C$th row of $A$ is the flattened upper triangular part of matrix $M_C$. In the demonstration, we can compare the result from 1D PCA projecting this data matrix to $\tilde{A}\in\mathbb{R}^{39\times 4}$ and use the row vector $\left(\tilde{A}^{(C)} \right)^\top\in\mathbb{R}^4$ associated to country $C$ in the clustering.

\subsection{Clustering algorithm}

Grouping of data points has become significant in various fields such as health, medicine, social, and spatial \cite{RU}. Solving this problem requires, for instance, a clustering technique to identify groups within a data set such that the similarity within the cluster is high and is low in-between \cite{AK}. Notably, there are several methods for clustering such as distance-based, connectivity-based methods, density-based and probabilistic techniques \cite{AK, BB, HF}.

Since we have a very small data set (i.e. number of data points is very low), the approaches considering dense set of data points do not work well now. In this study, agglomerative hierarchical clustering algorithm has been used due to its wide range of applications, simplicity and ease of implementation relative to other clustering methods. This technique is expected to give better results in comparison to other methods, where large amount of data is considered. In hierarchical clustering, the closest sets of clusters are merged at each level using a linkage method and then the dissimilarity matrix is updated correspondingly \cite{AK, CC, ML, NB, RU, SC}. This iterative process continues until only one maximal cluster remains. 

Several proximity measures are available for this clustering technique, e.g. single, complete, average, and ward's criterion. We have used complete linkage method as it takes the cluster structure into consideration (non-local in behavior) and generally obtains compact shaped clusters \cite{AK, ML}. For any clusters $C_{i}$ and $C_{j}$, this linkage measures the similarity of clusters as the similarity of their most dissimilar members, that is, it considers the longest distance among all their observations:
\begin{equation*}
     D_{\mathrm{complete}}(C_{i}, C{_j}) = \max \{d(x, y): x \in C_{i}, y \in C_{j}\} 
\end{equation*}
where $d$ is a chosen distance measure. In this work, we have considered the Euclidean distance, but other $L_p$ norms or specialized distance metrics can be defined (in many applications, the distance metric is the key to a good clustering).

Using the dissimilarity matrix and the linkage method, we can construct a dendrogram showing the agglomeration process during the running of the algorithm. The optimal number of clusters is obtained by manually cutting the tree at any given level and obtaining the clusters correspondingly without re-running the algorithm. The rule of thumb we applied, considers the longest vertical distance without any horizontal line passing through it. i. e. cutting the dendrogram where we have the largest gap between two successive merges \cite{AK}.

\begin{figure}[ht]
    \centering
    \includegraphics[width=1\textwidth]{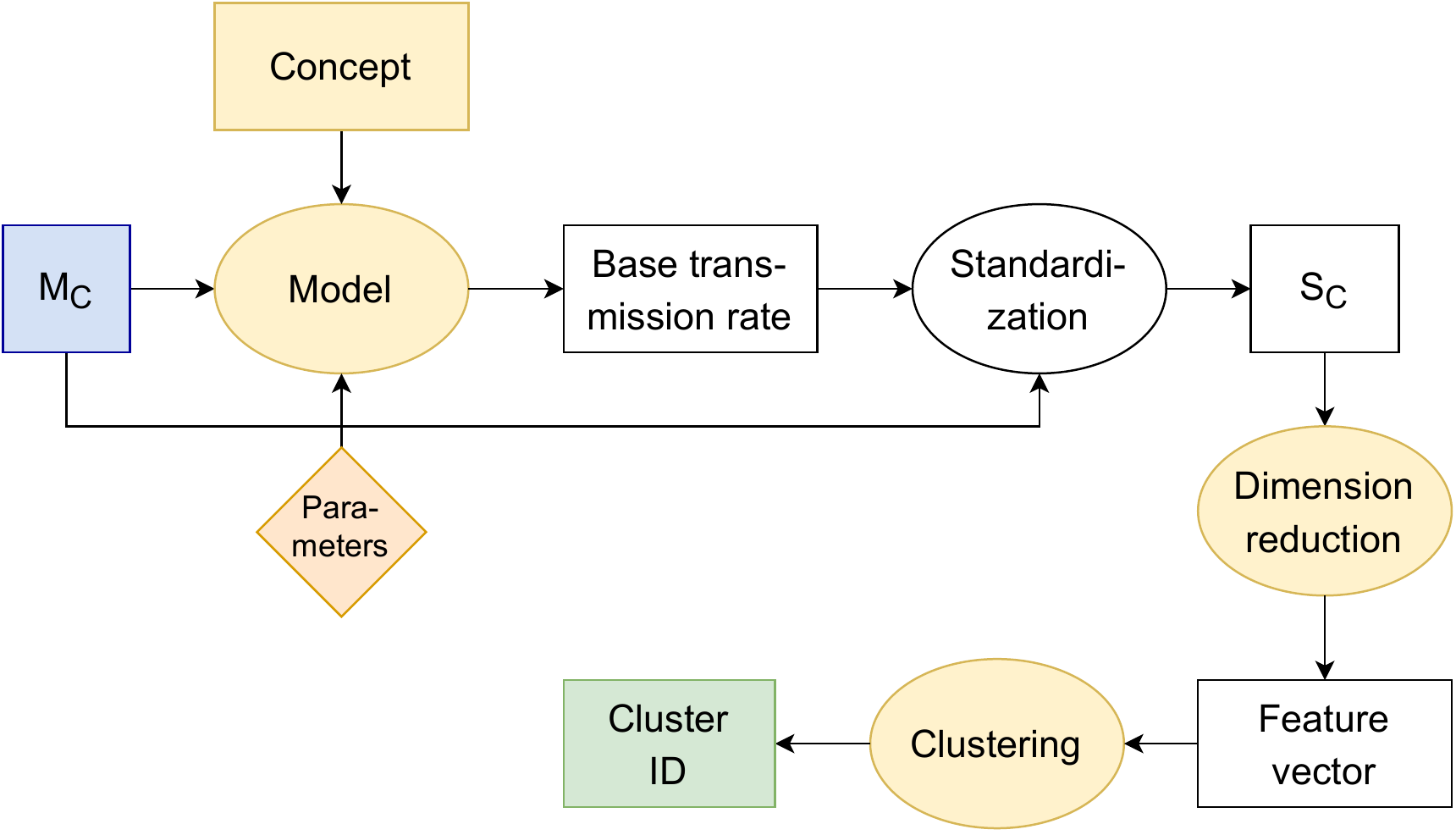}
    \caption{Flowchart showing a summary of the proposed framework. Rectangles show the output of a module depicted by ellipsoids, which contain function calls executing a specific part of the pipeline. The yellow shapes are generic and can be replaced by considering improved/application-specific alternatives. As it can be seen, the input of the procedure is the contact matrices and cluster correspondence computed at the output. The concept associated with the model describes along which indicator the standardization has to be done.}   
    \label{fig:pipeline}
\end{figure}

\subsection{The proposed framework}

The complete pipeline of the framework can be seen in Fig.\ref{fig:pipeline}.  
The main input of the pipeline is the list of country-specific contact matrices (or any list of contact matrices for which the grouping needs to be executed). The symmetrization of the matrices is done at loading the data to ensure consistent total number of contacts between the age groups. The resulted matrices inputted to an epidemic model, which is specified by the user/domain expert and then the base transmission rate $\beta_0$ is calculated along a concept. The concept determines the indicator for the calculation of $\beta_0$: in this paper, we use $\mathcal{R}_0$ to calculate this parameter, but a viable concept would be to obtain that $\beta_0$ for which the final epidemic size or final number of deaths is a pre-defined fraction of the population (this can be done for all countries). Using the symmetrized matrices and the previously calculated $\beta_0$s, we apply the standardization for the matrices via scaling the elements of them. 

In the next phase of the procedure we apply the data science-related steps on the standardized matrices. First a dimension reduction is applied to avoid curse of dimensionality in the next steps. The outputted low-dimensional vectors (i.e. the feature vectors) are used in the clustering algorithm, which is preferred to be a connectivity-based approach for very small datasets. In Fig. \ref{fig:pipeline}, yellow colored shapes show the generic parts of the pipeline (i.e. these parts are determined by the expert specifying the choices based on the needs and goals). The rectangular shapes gives the output data from the modules depicted by ellipsoids.

The proposed framework is implemented in Python using object-oriented programming to keep the code generic and easily extendable. We used open source packages to solve differential equations, eigenvalue problems and execute dimensionality reduction and clustering. The code is publicly available in Github \cite{ZE}.

\section{Results}

In this section, we demonstrate the previously introduced framework on the age-structured model of \cite{RB} with conceptually simple and correct algorithms for dimensionality reduction and clustering. Since \cite{PK} offers contact matrices for 16 age groups, we aligned the age-dependent parameters of the model. Here we investigate a total of 39 countries in Europe (see Appendix A.) to avoid significant differences between the countries from viewpoint of economics, location, demographics and culture. Following the methodology, both 1D PCA and $(\mathrm{2D})^{2}$ PCA approaches were applied to project into $\mathbb{R}^4$ to highlight importance of the proper reduction for matrix-valued data. The clustering algorithm was chosen to be an connectivity-based one with agglomerative execution and using Euclidean distance and complete linkage for merging the clusters during the procedure. For the simulation we chose $\mathcal{R}_0 = 2.2$ for calculating country-specific transmission rates.

\begin{figure}
    \centering
    \includegraphics[width=0.49\textwidth]{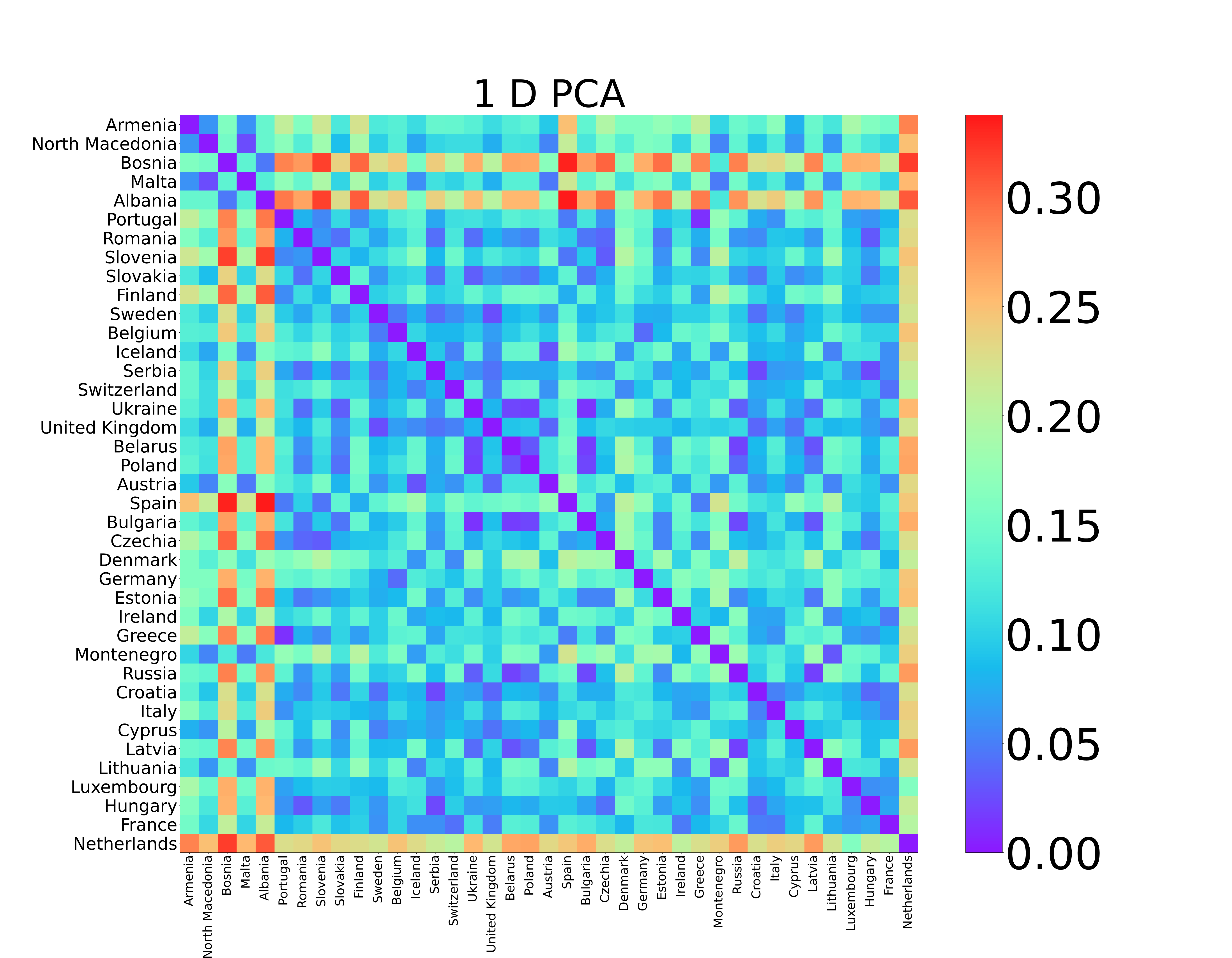}
    \includegraphics[width=0.49\textwidth]{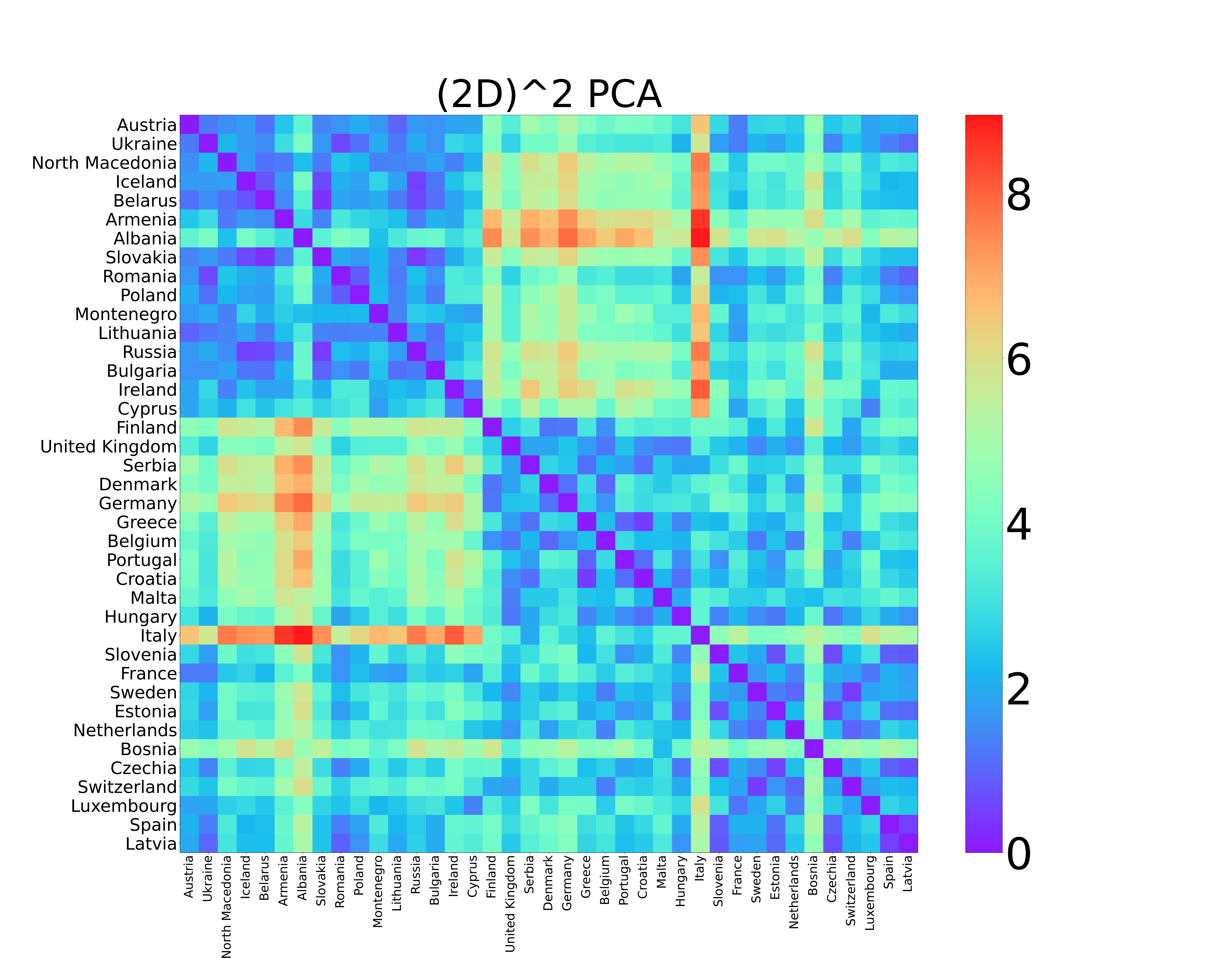}
    \caption{The pairwise distance matrices computed for all countries based on the feature vector from the dimensionality reduction step. The left figure corresponds to 1D PCA while the right is calculated from the result of the $(\mathrm{2D})^{2}$ PCA approach. Here we used Euclidean distance during the computations. We applied the dendrogram from the hierarchical clustering to arrange the rows and columns of the matrices providing a nicer presentation of the elements.}
    \label{fig: distances}
\end{figure}

Applying standardization for the matrices and executing dimensionality reduction, the computed feature vectors can be used to calculate pairwise Euclidean distance for the countries. For a nicer representation, the rows and columns of the distance matrix can be permuted using the dendrogram from the hierarchical clustering. The left figure in \ref{fig: distances} shows the distance matrix computed from the result of 1D PCA: as it can be seen, countries with dark blue color entries such as Bulgaria, Latvia and Belarus have very close social mixing patterns and such countries are expected to belong to the same cluster. Notably, countries such as Netherlands, Albania and Bosnia have orange and red cell colors respectively with most countries indicating somehow different social contact patterns. Such countries can have different cluster and seemingly different intervention strategies. The same intuition applies to $(\mathrm{2D})^{2}$ PCA approach in Fig. \ref{fig: distances}.

\begin{sidewaysfigure}
  \centering
    \includegraphics[width=\textheight]{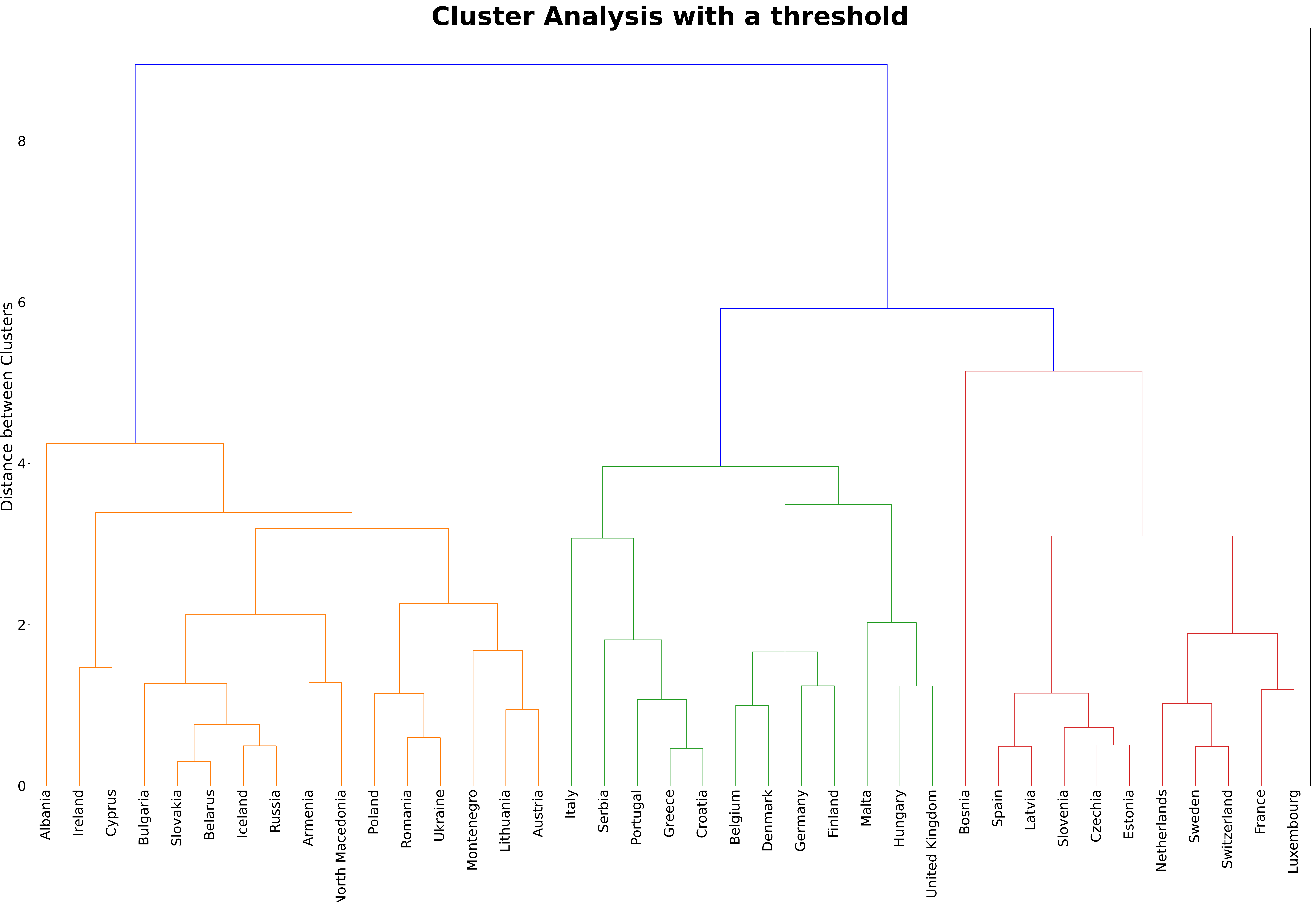}
    \caption{Dendrogram associated with the clustering of countries using feature vector generated by $(\mathrm{2D})^{2}$ PCA. Vertical axis indicates the distance between the two clusters being connected. The agglomerative manner of the algorithm can be read from following the tree from bottom to top.}
  \label{fig: dpca dendo}
\end{sidewaysfigure}

The advantage of using hierarchical clustering method is that it allows for manually cutting the hierarchy at any given level and obtaining the clusters correspondingly just by defining a threshold. The choice of threshold depends completely on the user. Dividing the dendrogram from 1D PCA at a cluster distance of 0.25, three separate clusters were formed, Netherlands became a single cluster. In the same way, cutting the dendrogram at a distance of 5.5 between the clusters calculated on reduced data from $(\mathrm{2D})^{2}$ PCA technique, the dendrogram formed three meaningful clusters (\ref{fig: dpca dendo}). The final list of clusters is provided in tables \ref{tblpca} and \ref{tbldpca} (for the former one, the one-element cluster with Netherlands was omitted from the table). Notice, that cluster sizes are highly imbalanced for 1D PCA, which seems to show the flaw of this reduction approach compared to the result using $(\mathrm{2D})^{2}$ PCA, where cluster sizes are more comparable, which indicates better separation after the projection.

\begin{sidewaysfigure}
  \centering
    \includegraphics[width=\textheight]{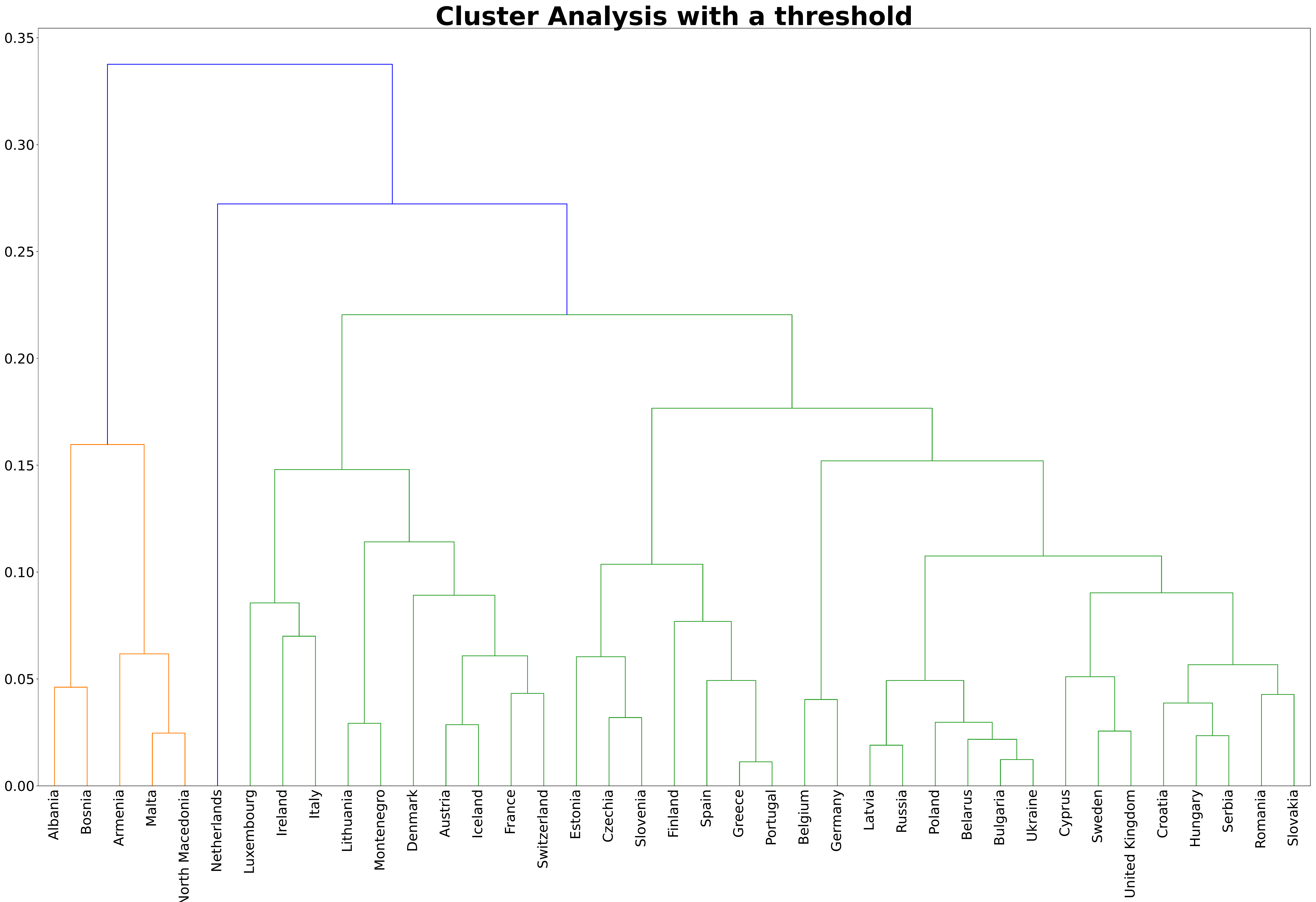}
    \caption{Dendrogram associated with the clustering of countries using feature vector generated by 1D PCA. Vertical axis indicates the distance between the two clusters being connected. The agglomerative manner of the algorithm can be read from following the tree from bottom to top.}
  \label{fig: pca dendo}
\end{sidewaysfigure}

\begin{table}[h]
{\footnotesize
\begin{tabular}{||p{3.6cm}|p{6.7cm}||}
\hline
{} & {} \\[-1.5ex]
Cluster 1 & Cluster 2\\[1ex]
\hline
Albania, Bosnia, Armenia, Malta, North Macedonia.
& Luxembourg, Ireland, Italy, Lithuania, Montenegro, Denmark,  Austria, Iceland, France, Switzerland, Estonia, Czechia, Slovenia, Finland, Spain, Greece, Portugal, Belgium,  Germany, Latvia,  Russia , Poland, Belarus, Bulgaria, Ukraine, Cyprus, Sweden, United Kingdom, Croatia, Hungary, Serbia, Romania, Slovakia. \\[1ex]
\hline
\end{tabular}
\caption{Clusters produced using 1D PCA. Netherlands is one element cluster and thus it is omitted in this listing. Clusters produced by the agglomerative hierarchical clustering algorithm based on social contact patterns for the 39 European countries applying projection 1D PCA. Cluster 1 and 2 have 5 and 33 elements, respectively. Netherlands is one element cluster and thus it is omitted in this listing.}
% \vspace*{-13pt}
\label{tblpca}}
\end{table}

\begin{table}[h]
{\footnotesize
\begin{tabular}{||p{3.4cm}|p{3cm} |p{3.4cm}||} 
\hline
{} & {} & {}\\[-1.5ex]
Cluster 1 & Cluster 2 & Cluster 3\\[1ex]
\hline
 Albania, Ireland, Cyprus, Bulgaria, Slovakia, Belarus, Iceland, Russia, Armenia, North Macedonia, Poland, Romania, Ukraine, Montenegro, Lithuania, Austria. & Italy, Serbia, Portugal, Greece, Croatia, Belgium, Denmark, Germany, Finland, Malta, Hungary, United Kingdom. & Bosnia, Spain, Latvia, Slovenia, Czechia, Estonia, Netherlands, Sweden, Switzerland, France, Luxembourg.\\[1ex]
\hline
\end{tabular}
\label{tbldpca}}
\caption{Clusters produced by the agglomerative hierarchical clustering algorithm based on social contact patterns for the 39 European countries applying projection $(\mathrm{2D})^{2} PCA)$. Cluster 1, 2 and 3 have 16, 12 and 11 elements, respectively}
\vspace*{-13pt}
\end{table}

\begin{figure}
    \centering
    \includegraphics[width=0.3\textwidth]{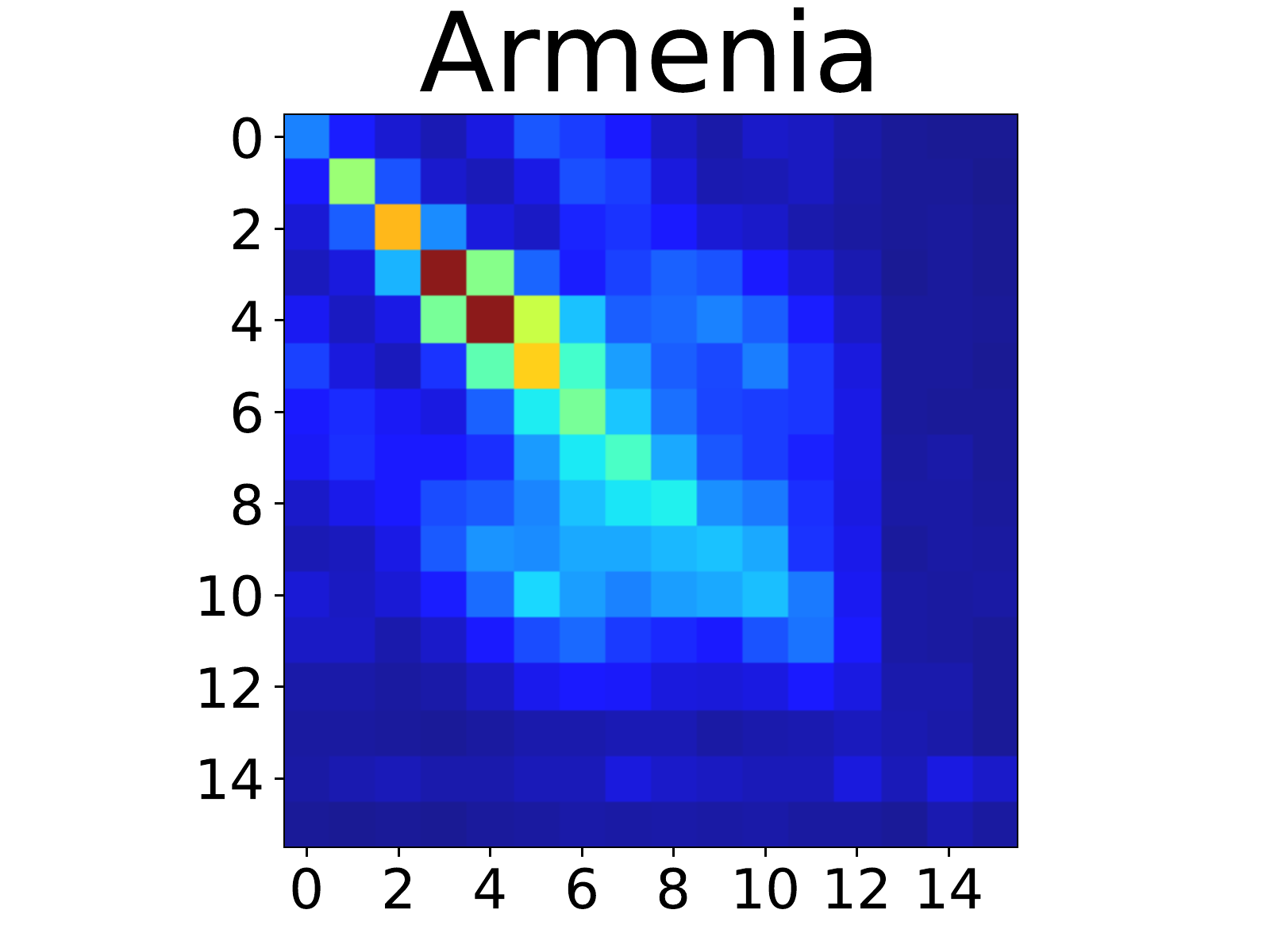}
    \includegraphics[width=0.3\textwidth]{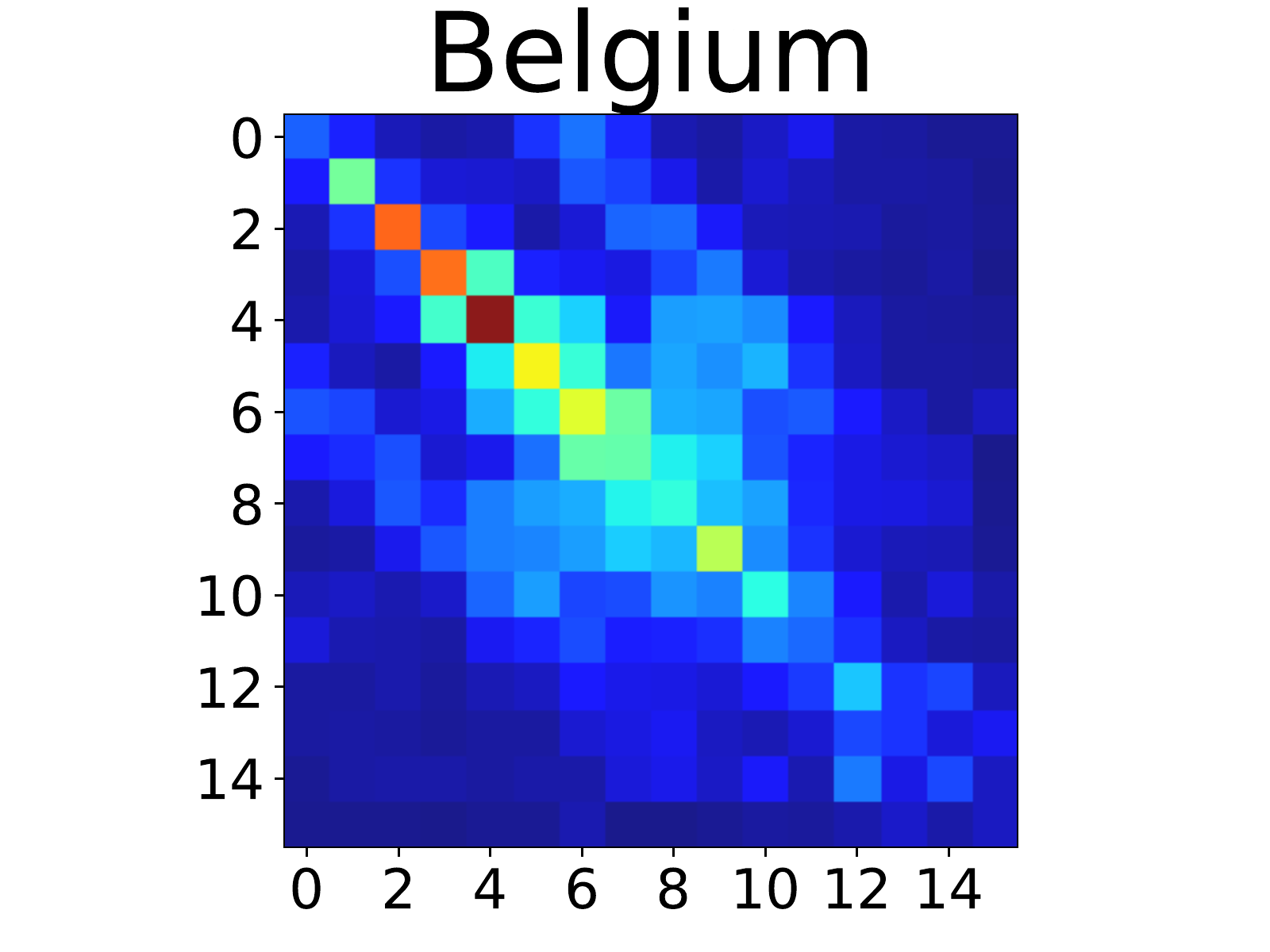}
    \includegraphics[width=0.3\textwidth]{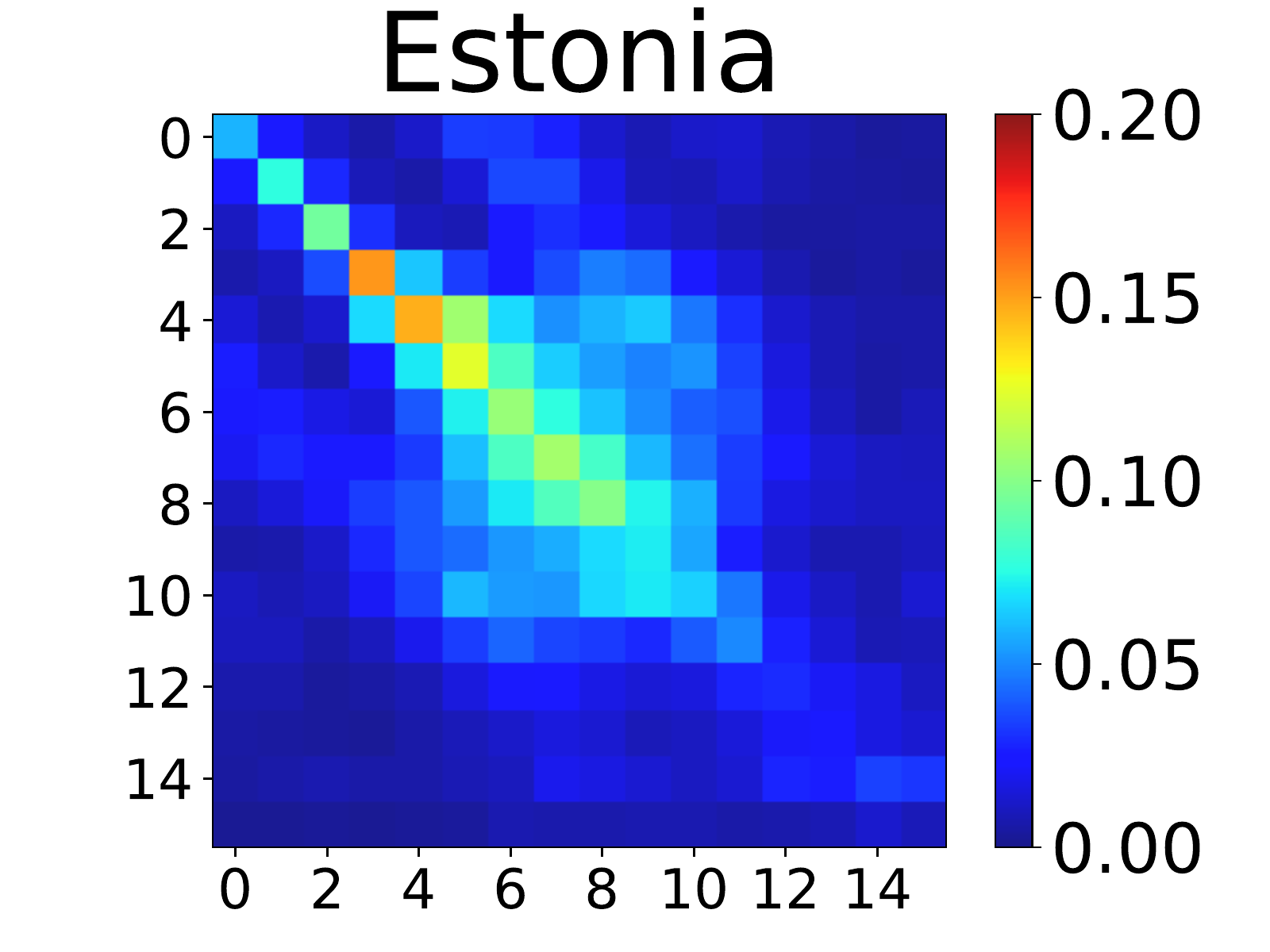}
    \caption{Standardized contact matrices correspond to Armenia (left), Belgium (centre), and Estonia (right), selected from different clusters resulted by clustering based on the feature vectors generated by the $(\mathrm{2D})^{2} PCA$ technique. These countries are located "in the middle" of the ordered list of the cluster elements, based on the dendrograms. Differences between these matrices appear along the elements in the main diagonal (larger number for younger and old age groups for Belgium), in the elements along diagonals paralel to the main one (expressing higher contact rates between "adult children" and older/retired parents) and middle square of the matrices (which come mainly from work contact).}
    \label{fig: contacts}
\end{figure}

Fig.\ref{fig: contacts} shows one element from each of the clusters calculated from reduced vectors resulted by $(\mathrm{2D})^{2}$ PCA. These countries (Armenia, Belgium and Estonia) were chosen, since that ones are located in the middle of the blocks along the horizontal axis. Considering these matrices, we see that differences occur between the matrices along the main diagonal (for Armenia, the elements for the older age groups are not as strong as the rest of the elements are not that large), along the diagonal parallel to the main diagonal (these express the contacts between the adult parents and retired grandparents, here Belgium and Estonia show more concentrated rates) and in the middle square of the matrix (this part comes mainly from workplace contacts, the distribution of the elements differ for all three of the samples). It is important to mention, that distribution of the elements is directly related to the statistical method described in \cite{PK}, since stronger smoothing effect can be recognized for estimation with less amount of data, thus distance measuring will is highly affected by the quality of the estimations. Nevertheless, the dimensionality reduction techniques developed for matrices aim to preserve the patterns in the structure of the matrices, for which the previously mentioned set of elements (diagonals, middle submatrix) give examples.

\section{Discussion}

Since emerging infectious diseases do not occur at the same time in all countries, it is important to understand, which countries are affected in the same way with a non-pharmeceutical intervention. The goverment of a country where the outbreak happens later in time can benefit from this information to make better decision in prevention and damage control. This paper proposes a framework which can support decision makers in determining, which other countries have to be primarily monitored and analyzed based on social mixing patterns. The underlying pipeline assumes only the contact matrices in the population and transforms it via an epidemic model and the concept of the analysis, then applies dimension reduction on this data and executes clustering on the projected data points associated with the considered countries. A realization of the framework was demonstrated with simple and correct approaches for the generic parts such as the epidemic model, the dimensionality reduction and the clustering.

In this procedure we can replace the concept of the analysis, since $\mathcal{R}_0$ describes only the transmission spread of a disease, but if a country primarily wants to minimize the number of fatal outcomes or the number of patients admitted to the hospital, these aspects can be also used for determining the country-specific scaling parameter in the above mentioned standardization. As an example, for all countries we can determine that $\beta_0$, for which the ratio of fatal outcomes is a fixed proportion. Clearly, the epidemic model can be replaced by any pre-defined model written for any infectious disease. For dimensionality reduction we applied a simple technique developed for matrices, but other approaches might be suitable for this part, see \cite{WS} for similar, but more sophisticated methods, but convolution-based techniques can be developed or modified optimization problem can be written down for the matrix reconstruction technique. For very small data, connectivity-based algorithms perform well and many different approaches are available (agglomerative and divisive types, different similarity measures and linkage methods). Globally, we can extend the feature vectors after the dimension reduction with other descriptors about the countries, which can be relevant in the analysis, e.g. some economical indicators, which can be relevant for the NPI strategy. Finally, the full contact matrix used as a main input to the procedure can also be computed with other weighting in the summation to give more importance to specified type of contacts, e.g. home contacts are really important, since typical NPIs reducing contacts (school closure, home office, closing malls) leave this contact unchanged (or even slightly increase them).

\section{Acknowledgement}
This research was supported by project TKP2021-NVA-09, implemented with the support provided by the Ministry of Innovation and Technology of Hungary from the National Research, Development and Innovation Fund, financed under the TKP2021-NVA funding scheme.

\appendix

\section{First Appendix}   

\subsection{List of the European countries}

We considered the following European countries in the demonstration:\\
Albania, Armenia, Austria, Belarus, Belgium, Bosnia, Bulgaria, Croatia, Cyprus, Czechia, Denmark, Estonia, Finland, France, Germany, Greece, Hungary, Iceland, Ireland, Italy, Latvia, Lithuania, Luxembourg, Malta, Montenegro, Netherlands, North Macedonia, Poland, Portugal, Romania, Russia, Serbia, Slovakia, Slovenia, Spain, Sweden, Switzerland, Ukraine, United Kingdom.

\section{Second Appendix}

\subsection{The Governing Equations of the Epidemic Model.}
The governing equations of the disease model described in Section 2.2 take the form

\begin{align}
 {S^i}'(t)={} & -\beta_0 \frac{S^i(t)}{N_i}\cdot\sum_{k=1}^{16} M^{(k,i)}\left[I_{p}^{k}(t) + \mathrm{inf}_a \sum_{j=1}^3 I_{a,j}^{k}(t) + \sum_{j=1}^3 I_{s,j}^{k}(t)\right] \nonumber\\ 
 {L_1^i}'(t)={} & \beta_0 \frac{S^i(t)}{N_i}\cdot\sum_{k=1}^{16} M^{(k,i)}\left[I_{p}^{k}(t) + \mathrm{inf}_a \sum_{j=1}^3 I_{a,j}^{k}(t) + \sum_{j=1}^3 I_{s,j}^{k}(t)\right] - 2 \alpha_l L^i_1(t) \nonumber\\
 {L_2^i}'(t)={} & 2 \alpha_l L_1^i(t) - 2\alpha_l L_2^i(t),\nonumber\\
 {I_a^i}'(t)={} & 2 \alpha_l L_{2}^{i}(t) - \alpha_{p} I_{p}^{i} (t)\nonumber\\
 {I_{a,1}^i}'(t)={} & p^{i} \alpha_{p} {I}_{p}^{i} (t) - 3 \gamma_{a} I_{a,1}^{i}(t)\nonumber\\
 {I_{a,2}^i}'(t)={} & 3\gamma_{a} I_{a,1}^{i}(t)- 3\gamma_{a} I_{a,2}^{i}(t)\nonumber\\ 
 {I_{a,3}^i}'(t)={} & 3\gamma_{a} I_{a,2}^{i}(t)- 3\gamma_{a} I_{a,3}^{i}(t)\\ 
 {I_{s,1}^i}'(t)= {}& (1 - p^i) \alpha_{p} I_{p}^{i} - 3 \gamma_{s} I_{s,1}^i(t)\nonumber\\
 {I_{s,2}^i}'(t)= {} & 3 \gamma_{s} I_{s,1}^i(t) - 3 \gamma_{s} I_{s,2}^i(t)\nonumber\\
 {I_{s,3}^i}'(t)= {} & 3 \gamma_{s} I_{s,2}^i(t) - 3 \gamma_{s} I_{s,3}^i(t)\nonumber\\
 {I_h^i}'(t)={} & h^i (1 - \xi^i) 3 \gamma_{s} I_{s,3}^i(t) - \gamma_h I_h^i(t)\nonumber\\
 {I_c^i}'(t)={} & h^i \xi^i 3 \gamma_{s} I_{s,3}^i(t)-\gamma_c I_c^i(t)\nonumber\\
 {I_{\mathrm{cr}}^i}'(t)={} &  (1 - \mu^{i}) \gamma_{c} I_{c}^{i} (t) -\gamma_{\mathrm{cr}} I_{\mathrm{cr}}^{i} (t)\nonumber\\
 {R^i}'(t)={} & 3 \gamma_{a} I_{a,3}^{i} (t) + (1- h^{i}) 3 \gamma_{s} I_{s,3}^{i} (t) + \gamma_{h} I_{h}^{i} (t) + \gamma_{\mathrm{cr}} I_{\mathrm{cr}}^{i} (t)\nonumber\\   
 {D^i}'(t)={} & \mu^i \gamma_c I_c^i(t),\nonumber
 \label{eq:model}
\end{align}
where the index $i \in {1, . . . ,16}$ represents the corresponding age group. As it can be seen, the model contains age-dependent parameters (probabilities $p$, $h$, $\xi$, $\mu$, for which upper index shows the age group) and age-independent ones (fraction $\mathrm{inf}_a$ and transition parameters $\alpha$ and $\gamma_X$, where $X\in\{a,s,h,c,\mathrm{cr}\}$). Notation here is aligned with the parameter file located in the repository of the framework. Here $\mathrm{inf}_a$ denotes the relative infectiousness of $I_a$ compared to $I_s$, for more details about the other parameters and the methodology for parametrization, see \cite{RB}.

\subsection{Next Generation Matrix}

To calculate $\mathcal{R}_{0}$ for the previously mentioned epidemic model, we consider the infectious subsystem for 
$${L_{1}^{i}}(t), {L_{2}^{i}}(t), {I_{p}^{i}}(t), {I_{j}^{i}}(t),$$ with
$j\in \boldsymbol\{a, s\} \times \{1,2,3\}, i \in \{1,...,16\}$, thus 
\begin{align*}
X(t) &= 
    \begin{bmatrix}
           L_{1}^{i}(t) & 
           L_{2}^{i}(t) & I_{p}^{i}(t) &
           I_{a, 1}^{i}(t) & 
           I_{a, 2}^{i}(t) & 
           I_{a, 3}^{i}(t) &
           I_{s, 1}^{i}(t) &
           I_{s, 2}^{i}(t)&
           I_{s, 3}^{i}(t)    
    \end{bmatrix}^\top
\end{align*} and linearization gives
$${X^\prime}(t) = (\beta_0 \cdot T + \Sigma) \cdot X(t),$$ 
where $T \in \mathbb{R}^{144 \times 144}$ is the transmission part and $\Sigma\in \mathbb{R}^{144 \times 144}$ represents the transition mechanisms in the model. The matrix $\Sigma$ is a block-diagonal matrix, where blocks have size of $9\times9$ containing transition parameters related to the linear terms of the system. On the other hand, the transmission matrix $T$ is partitioned into blocks of size $9\times9$ and each of this blocks have non-zero elements only in their first rows, since transmission between individuals affects only the classes $L_1^i, i\in\{1,2,\dots,16\}$ and these non-zero elements are related to the corresponding elements of the contact matrix and the transmission-related parameter $\mathrm{inf}_a$. The Next Generation Matrix (shortly NGM) can be calculated as $$\mathrm{NGM} = -\beta_0\cdot T \Sigma^{-1},$$ and the basic reproduction number is the dominant eigenvalue of the NGM, i.e. $$\mathcal{R}_0 = \beta_0 \cdot \rho(-T  \Sigma^{-1}).$$ On one hand, the model parameters are the same for all countries in the paper, on the other hand this calculation has to be executed for each countries separately, since social contact matrices (thus $T$ matrices) are different. For more details about NGM method, see \cite{DH}.

\end{document}